\documentclass[twocolumn]{jpsj2}
\title{%,0
Hole Doping Effects on Spin-gapped Na$_{2}$Cu$_{2}$TeO$_{6}$ via Topochemical Na Deficiency 
}

\author{%
Kumiko \textsc{Morimoto}$^{1}$, 
Yutaka \textsc{Itoh}$^{1}$\thanks{E-mail address: itoh@kuchem.kyoto-u.ac.jp},  
Kazuyoshi \textsc{Yoshimura}$^{1}$, Masaki \textsc{Kato}$^{2}$ and Ken \textsc{Hirota}$^{2}$
}

\inst{%
$^{1}$Department of Chemistry, Graduate School of Science,
Kyoto University, Kyoto 606-8502\\ 
$^{2}$Department of Molecular Science and Technology, 
Faculty of Engineering, Doshisha University,
Kyotanabe, Kyoto 610-0394, Japan\\ 
}

\recdate{\today}

\abst{%
We report the magnetic susceptibility and NMR studies of a spin-gapped layered compound
 Na$_{2}$Cu$_{2}$TeO$_{6}$ (the spin gap $\Delta\sim$ 250 K),  
 the hole doping effect on the Cu$_{2}$TeO$_{6}$ plane  via a topochemical Na deficiency 
 by soft chemical treatment,
 and  the static spin vacancy effect by nonmagnetic impurity Zn substitution for Cu. 
 A finite Knight shift at the $^{125}$Te site was observed for pure Na$_{2}$Cu$_{2}$TeO$_{6}$. 
 The negative hyperfine coupling constant $^{125}A_{\mathrm{tr}}$ is an evidence for the existence of 
a superexchange pathway of the Cu-O-Te-O-Cu bond. 
It turned out that 
both the Na deficiency and Zn impurities induce a Curie-type magnetism in the uniform spin
susceptibility  in an external magnetic field of 1 T, 
but only the Zn impurities enhance the  low-temperature $^{23}$Na nuclear spin-lattice relaxation rate
whereas the Na deficiency suppresses it.  
A spin glass behavior was observed for the Na-deficient samples but not for the Zn-substituted samples.
The dynamics of the unpaired moments of the doped holes are different from that of the spin vacancy
in the spin-gapped Cu$_{2}$TeO$_{6}$ planes.         
}

\kword{%
spin gap, Na$_{2}$Cu$_{2}$TeO$_{6}$, soft chemistry, hole doping, NMR
}

\begin{document}
\maketitle
  
Na$_{2}$Cu$_{2}$TeO$_{6}$ 
is a  spin-gapped layered compound and a Mott insulator~\cite{Xu}. 
Non-orthogonal spin dimers lie on Cu$_{2}$TeO$_{6}$ layers piled
up with Na layers alternately. 
The uniform magnetic susceptibility $\chi$ shows 
a broad maximum at about $T_{\mathrm{max}}$ = 160 K
and a spin gap $\Delta\sim$ 250 K ($\Delta$ was estimated from Ref. [1]). 
Although the crystal structure is two-dimensional,
the temperature dependence of $\chi$ indicates an alternating exchange Heisenberg chain. 
The large spin gap of about 250 K indicates the existence of a strong
superexchange interaction on the Cu$_{2}$TeO$_{6}$ layers. 
 Figure 1 illustrates the Cu$_2$TeO$_6$ plane of Na$_{2}$Cu$_{2}$TeO$_{6}$
 and the $J_{1}$-$J_{2}$-$J_{3}$ superexchange pathways on the plane. 
 According to Ref. [1], the alternating exchange interactions are
predominant  $J_{1}$ and the weaker $J_{2}$ = 0.3$J_{1}$.   
We address the following questions: whether the relatively long Cu-O-Te-O-Cu path
actually bears such a strong superexchange interaction, 
and what kind of magnetic instability is behind the
magnetic excitation spectrum with the large spin gap. 
The dynamics of the doped holes on the gapped magnetic excitation spectrum 
has attracted much attention 
as well as the hole-doped N{\'e}el ordering state~\cite{Machi,Ishida}. 
 
     In this paper, we studied the low-lying excitations of pure Na$_{2}$Cu$_{2}$TeO$_{6}$, 
the hole doping effect via a Na deficiency, 
and the spin vacancy effect via the Zn substitution of the Cu sites  
by means of magnetic susceptibility and NMR measurements.  
We obtained evidence of the existence of small covalency at the Te site
by the $^{125}$Te NMR technique. 
We succeeded in synthesizing the topotactic Na-deficient samples by soft chemical treatment, 
i.e., by a redox reaction with an oxidizer and the mother compound Na$_{2}$Cu$_{2}$TeO$_{6}$. 
The Na deficiency and the Zn substitution were found to induce Curie-type terms in their magnetic
susceptibilities at low temperatures. 
No impurity-induced magnetic long range ordering was confirmed
for both the cases down to 2 K. 
For the Na-deficient samples, however, a spin-glass behavior was
observed in the spin-gapped state at low temperatures, at the maximum of $T_{g}\sim$ 15 K. 
The doped holes via the Na deficiency were found to induce competition
among the interactions on the Cu$_{2}$TeO$_{6}$ layers. 

Pure Na$_{2}$Cu$_{2}$TeO$_{6}$ samples were synthesized by a conventional solid state reaction method. 
The starting materials $-$Na$_{2}$CO$_{3}$, CuO, and TeO$_{2}$$-$ were ground in a molar ratio
of 1:2:1, pelletized, and heated at 680 $^\circ$C for 36 h in flowing oxygen gas. 
The resultant products were ground, repelletized, and heated under the same conditions.
Zn-substituted samples were also synthesized by the solid state reaction method
with ZnO as a starting material.  

\begin{figure}[h]
\begin{center}
\includegraphics[width=7.3 cm]{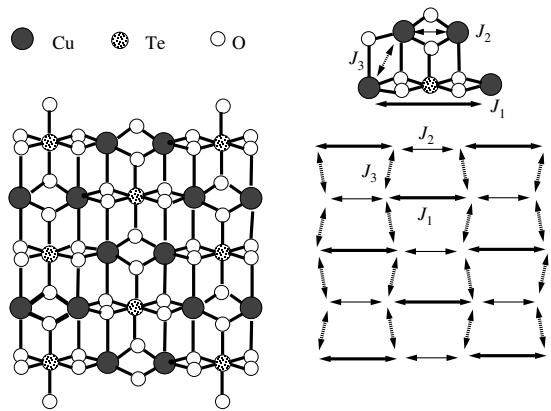}
\end{center}
\caption{%
Top view of a Cu$_{2}$TeO$_{6}$ plane of Na$_{2}$Cu$_{2}$TeO$_{6}$ (left panel).
A schematic $J_{1}$-$J_{2}$-$J_{3}$ model illustrates   
the superexchange pathways according to Ref. [1].
}
\label{Structure}
\end{figure}  
 
Na-deficient samples Na$_{2-\delta}$Cu$_{2}$TeO$_{6}$ were prepared by the soft chemical oxidation of the mother compound Na$_{2}$Cu$_{2}$TeO$_{6}$, 
using various concentrations of bromine as the oxidizing agent,
similar to Na$_{x}$CoO$_{2}$.~\cite{Cava} 
Na$_{2}$Cu$_{2}$TeO$_{6}$ powder, typically 0.4 g, was stirred in 5 ml of a bromine solution 
in acetonitrile at 25 $^\circ$C for 18 h.
The products were washed several times with water and then with
acetone to remove bromine and dried briefly under an ambient condition. 

As the bromine concentration $y$ increases 
as 1.0, 1.5, 1.6, 2.0, 2.5, 5.0, and 10 mol/L of Br$_{2}$/MeCN,
the amount of Na deficiency increases,
which was confirmed by inductively coupled plasma atomic emisson spectroscopy (ICP-AES).
The typical value of the ratio of Na to Cu was estimated
to be (2 - $\delta$)/2 = 0.93 (1.5 M) and 0.80 (2.5 M)
within our technical accuracy of the ICP-AES. 
Because of the charge neutrality,
the Na deficiency $\delta$ is compensated by the holes in the Cu$_{2}$TeO$_{6}$ plane;
the nominal valence of Cu is then changed 
from 2 in the stoichiometric compound into 2 + $\delta$/2. 
Thus, the nominal effective number $p$ of unpaired moments per Cu was estimated 
to be $p$ = 0.07 (1.5 M) and 0.20 (2.5 M). 
 
Magnetic susceptibility was measured by a superconducting quantum interference device (SQUID) magnetometer. 
$^{65}$Cu (nuclear spin $I$ = 3/2), $^{23}$Na ($I$ = 3/2) and $^{125}$Te ($I$ = 1/2) NMR were performed 
by a pulsed NMR technique. 
  
\begin{figure}[h]
\begin{center}
\includegraphics[width=7.3 cm]{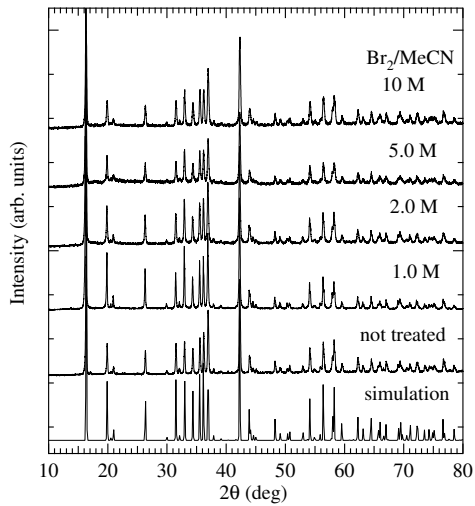}
\end{center}
\caption{%
Powder XRD patterns of the samples treated with Br$_{2}$/MeCN.
}
\label{XRD}
\end{figure} 
  
\begin{figure}[h]
\begin{center}
\includegraphics[width=7.0 cm]{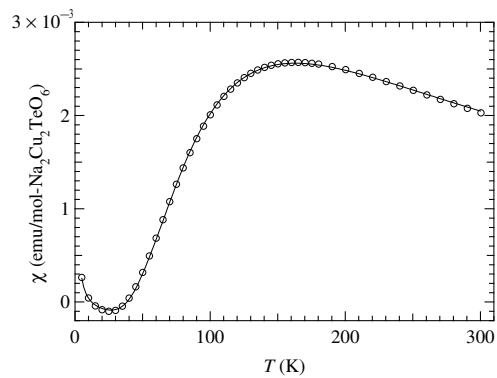}
\end{center}
\caption{%
Magnetic susceptibility of pure Na$_{2}$Cu$_{2}$TeO$_{6}$ at $H_{0}$ = 1 T.
The solid curve is the least squares fit result by eq. (1). 
}
\label{XPure}
\end{figure}
  
Figure 2 shows the powder X-ray diffraction (XRD) patterns of pure and Na-deficient samples. 
The powder XRD patterns show that the samples are in a single phase.  
As the bromine concentration increases, the individual lines of the XRD pattern
are broadened and tend to shift to the lower angle side within the broad linewidth, 
indicating the effect of the nonstoichiometry of the Na deficiency. 
 
Figure 3 shows the bulk magnetic susceptibility $\chi$ of pure Na$_{2}$Cu$_{2}$TeO$_{6}$.
The magnetic susceptibility $\chi$ per formula unit is expressed by
\begin{equation}
\chi=2\chi_{s}+\chi_{\mathrm{0}}+2C/T,
\label{e.TeKnightShift}
\end{equation}
 where  $\chi_{s}$ is the spin susceptibility, $\chi_{0}$ is the temperature-independent magnetic susceptibility and should be the sum of the Van Vleck orbital susceptibility 2$\chi_{\mathrm{orb}}$ and the core-electron diamagnetic susceptibility $\chi_{\mathrm{dia}}$ (= $-$0.116 $\times$ 10$^{-3}$ emu/mol-f.u.), and $C/T$ is a Curie term. 
The numerical factor 2 represents two Cu ions per formula unit. 
We assumed the theoretical eq. (56a) in Ref. [5] 
for the $S$ = 1/2 antiferromagnetic alternating-exchange Heisenberg chain 
as the spin susceptibility. 
Thus, $\chi_{s}$ is a function of the $g$-factor $g$, $J_{1}$, and 
an alternation parameter $\alpha$ ($\equiv$ $J_{2}$/$J_{1}$). 
The solid curve in Fig. 3 is the least squares fit result by eq. (1), 
where $g$, $J_{1}$, $\alpha$,  $\chi_{0}$, and $C$ are the fit parameters. 
We obtained
$g$ = 2.23, $J_{1}$ = 262 K,  $\alpha$ = 0.230,  $\chi_{0}$ = $-$0.179 $\times$ 10$^{-3}$ emu/mol-f.u.,
and $C$ = 1.09 $\times$ 10$^{-3}$ emu$\cdot$K/mol-f.u., which are consistent with the result in Ref. [1]. 
After Ref. [5], we obtained $\Delta$ = 243 K from $J_{1}$ and $\alpha$. 
The estimated $\chi_{0}$ smaller than the calculated $\chi_{\mathrm{dia}}$ might be
due to the remnants of unreacted materials beyond our ICP-AES analysis. 
 
\begin{figure}[h]
\begin{center}
\includegraphics[width=7.3 cm]{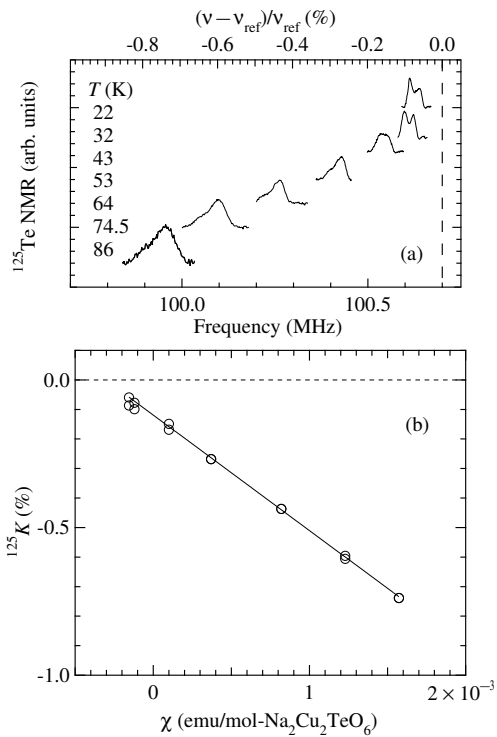}
\end{center}
\caption{%
(a) Frequency-swept $^{125}$Te NMR spectra of pure Na$_{2}$Cu$_{2}$TeO$_{6}$
at $H_{0}$ = 7.4847 T
below the peak temperature 160 K of the magnetic susceptibility $\chi$. 
The dashed line is the reference resonance position $\nu_{\mathrm{ref}}$ = 100.70 MHz
at $H_{0}$. The nuclear gyromagnetic ratio $^{125}\gamma_{n}$ is assumed to be 13.454 MHz/T. 
The top axis represents the frequnecy shift defined by ($\nu$ $-$ $\nu_{\mathrm{ref}}$)/$\nu_{\mathrm{ref}}$. 
A slight anisotropic powder pattern might result from the 5$s^{2}$ lone pair of Te$^{4+}$ ions. 
(b)  $^{125}$Te Knight shift $^{125}K$ plotted against magnetic susceptiblity $\chi$ with temperature as an implicit parameter ($K$-$\chi$ plot). 
}
\label{TeNMR}
\end{figure}
 
Figure 4(a) shows the frequency-swept $^{125}$Te NMR spectra of pure Na$_{2}$Cu$_{2}$TeO$_{6}$
in an external magnetic field of $H_{0}$ = 7.48473 T
below $T_{\mathrm{max}}$ = 160 K. 
Figure 4(b) shows the $^{125}$Te Knight shift $^{125}K$ plotted against 
the magnetic susceptibility $\chi$ with temperature as an implicit parameter.
The linear relation in the $K$-$\chi$ plot indicates that the spin Knight shift at the Te site
results from the local magnetic polarization by the Cu electron spins
 
\begin{equation}
^{125}K_{s}= {^{125}A_{\mathrm{tr}}}\chi_{s},
\label{e.TeKnightShift}
\end{equation}
where $^{125}A_{\mathrm{tr}}$ is a hyperfine coupling constant.
  
 From Fig. 4(b), we estimated 
the hyperfine coupling constant of 
$^{125}A_{\mathrm{tr}}$ = $-$43.7 kOe/$\mu_{\mathrm{B}}$.
Since $^{125}A_{\mathrm{tr}}$ is nearly istropic,
the dipole field from Cu moments is not predominant  at the Te site. 
An exchange polarization transfer mechanism to an empty orbital of the metal ion
can yield such a negative hyperfine coupling.~\cite{Owen} 
Thus, the $J_{1}$-$J_{2}$ alternating superexchange chains in Fig. 1 can 
be a  canonical model of Na$_{2}$Cu$_{2}$TeO$_{6}$.  
 
 \begin{figure}[h]
\begin{center}
\includegraphics[width=6.8 cm]{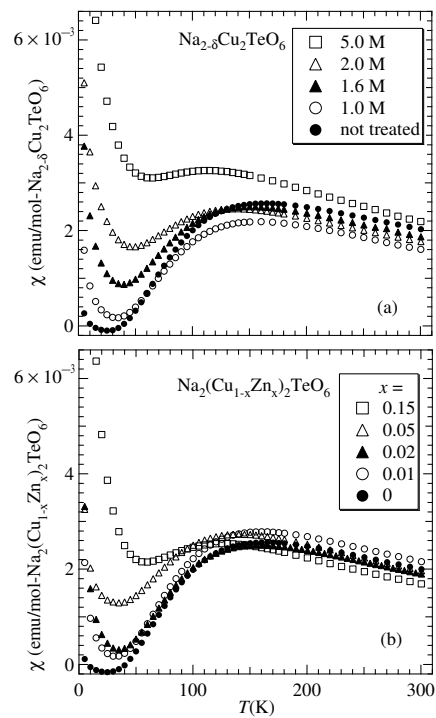}
\end{center}
\caption{%
(a) Magnetic susceptiblity $\chi$ of Na$_{2-\delta}$Cu$_{2}$TeO$_{6}$ 
treated with $y$ mol/L of Br$_{2}$/MeCN 
($y$ = 1.0, 1.6, 2.0, and 5.0) measured by the SQUID magnetometer at $H_{0}$ = 1 T. 
(b) Magnetic susceptiblity $\chi$ of Na$_{2}$(Cu$_{1-x}$Zn$_{x}$)$_{2}$TeO$_{6}$ with $x$ = 0.01, 0.02, 0.05, and 0.15 at 1 T. 
Both Na deficiency and Zn substitution cause a Curie magnetism at low temperatures. 
}
\label{XT}
\end{figure}
\begin{figure}[h]
\begin{center}
\includegraphics[width=6.8 cm]{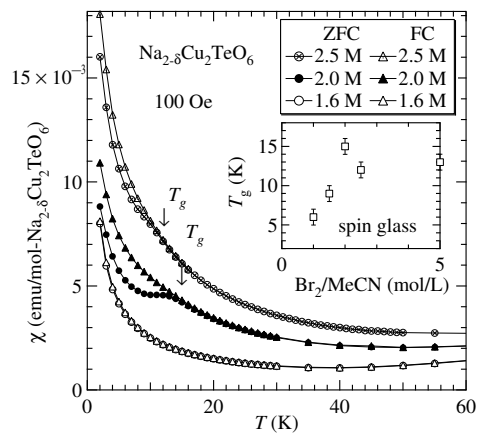}
\end{center}
\caption{%
Low-field hysterisis of magnetic susceptibility $\chi$ of Na$_{2-\delta}$Cu$_{2}$TeO$_{6}$ treated with $y$ mol/L of Br$_{2}$/MeCN 
($y$ = 1.6, 2.0, and 2.5) after cooling in a zero magnetic field (ZFC) and an external magnetic field of 100 Oe (FC). 
Inset shows the spin glass temperature $T_{g}$ against the bromine concentration $y$ M. 
}
\label{SG}
\end{figure}

Figures 5(a) and 5(b) show the effects of Na-deficiency and Zn-substitution on
magnetic susceptibility $\chi$ 
for Na$_{2-\delta}$Cu$_{2}$TeO$_{6}$ treated with $y$ mol/L of Br$_{2}$/MeCN 
($y$ = 1.0, 1.6, 2.0, and 5.0) and 
for Na$_{2}$(Cu$_{1-x}$Zn$_{x}$)$_{2}$TeO$_{6}$ with $x$ = 0.01, 0.02, 0.05, and 0.15
at $H_{0}$ = 1 T, respectively.
Both the Na deficiency and Zn impurities cause the Curie susceptibility at low temperatures. 
The Curie susceptibility induced by the Na deficiency in Fig. 5(a)
is similar to the Na-deficient effect on Na$_{x}$V$_{2}$O$_{5}$~\cite{Isobe}. 

We fitted a Curie-Weiss law of $C/(T-\Theta)$ to reproduce the low-temperature $\chi$
in Fig. 5. 
For the Na-deficient samples, $|\Theta|$ increased 
as the bromine concentration increased, 
e.g., $\Theta\sim$ $-$2.3 K for 1.6 M and $-$5.0 K for 2.0 M,
whereas 
for the Zn-substituted samples, $\Theta$ was nearly zero.
The results indicate that 
the doped holes are coupled with each other in the spin-gapped state,
whereas
the Zn-induced moments are nearly free. 
The Curie constants $C$ for the Na-deficient samples were smaller than 
what would be expected from the effective hole number $p$ in the ICP-AES analysis.
For the Na-deficient and Zn-substituted samples with nearly the same Curie terms, 
the hole number $p$ was larger than the Zn concentration $x$. 
The doped holes may not be well localized at specific sites on the Cu$_{2}$TeO$_{6}$. 
For both the systems, the product of the maximum magnetic susceptibility $\chi_{\mathrm{max}}$
and the maximum temperature $T_{\mathrm{max}}$ slowly decreased 
as the Na deficiency and Zn impurity increased. 

\begin{figure}[h]
\begin{center}
\includegraphics[width=7.0 cm]{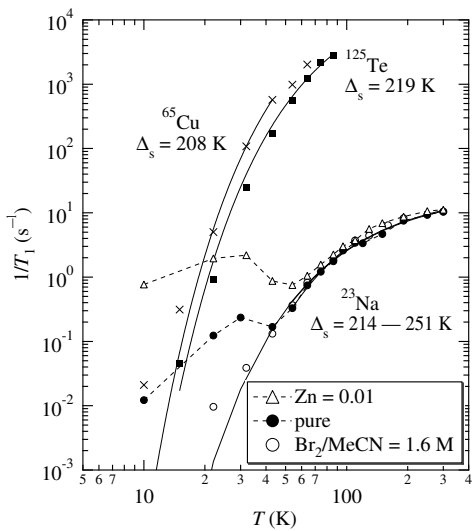}
\end{center}
\caption{%
Log-log plots of $^{23}$Na nuclear spin-lattice relaxation rate 1/$T_{1}$ against temperature
for pure Na$_{2}$Cu$_{2}$TeO$_{6}$ (closed circles), 
Na-deficient Na$_{2-\delta}$Cu$_{2}$TeO$_{6}$ treated with 1.6 mol/L of Br$_{2}$/MeCN (open circles),
and Zn-substituted Na$_{2}$(Cu$_{1-x}$Zn$_{x}$)$_{2}$TeO$_{6}$ with $x$ = 0.01 (open  triangles)
at $H_{0}$ = 7.4847 T. 
For pure Na$_{2}$Cu$_{2}$TeO$_{6}$, $^{63}$Cu (crosses) and $^{125}$Te (closed squares) nuclear spin-lattice relaxation rates
are also shown. 
The solid curves are the least squares fits by an activation function of exp($-\Delta_{s}/T$).
The estimated value of $\Delta_{s}$ slightly depends on the Cu, Te and Na sites and on the fit temperature ranges.
 }
\label{T1}
\end{figure}

Figure 6 shows the low-field hysteresis of the magnetic susceptiblity $\chi$ of Na$_{2-\delta}$Cu$_{2}$TeO$_{6}$ treated with $y$ mol/L of Br$_{2}$/MeCN 
($y$ = 1.6, 2.0, and 2.5), measured 
after cooling in a zero magnetic field (ZFC) and in a magnetic field of 100 Oe (FC). 
The magnetic hysteresis was easily suppressed by a high magnetic field.  
Thus, the bifurcation temperature may be regarded as a spin glass temperature $T_{g}$. 
For the sample with $y$ = 2.0 M, $T_{g}$ = 14 K up to $H_{0}$ = 20 Oe was suppressed 
by 1 T.   
The inset shows the spin glass temperature $T_{g}$ against the bromine concentration $y$ M. 
Such a spin glass behavior induced by doped holes 
is known for the hole-doped La$_{1.86}$Sr$_{0.04}$CuO$_{4}$
in a semiconducting regime.~\cite{Chou}
No hysteresis was observed for the Zn-substituted samples at 100 Oe. 
 
 In Fig. 7, 
 the activation behaviors of the nuclear spin-lattice relaxation rates 1/$T_{1}$ 
 of $^{23}$Na, $^{125}$Te, and $^{65}$Cu nuclear spins 
as a function of temperature were observed for Na$_{2}$Cu$_{2}$TeO$_{6}$, 
indicating the existence of a spin gap in the magnetic excitation spectrum.
From the fits by a function of exp($-\Delta_{s}/T$), we obtained the spin gap $\Delta_{s}$ = 208$-$251 K, 
which is nearly the same as the gap $\Delta\sim$ 250 K estimated from the static uniform spin
susceptibility. 
The value of $\Delta_{s}$ slightly depends on the nuclear sites and on the fit temperature ranges.

In Fig. 7, a small peak behavior of 1/$^{23}T_{1}$ in the spin-gapped state
was observed for Na$_{2}$Cu$_{2}$TeO$_{6}$ at about 30 K,
which has been frequently observed for quantum spin-gapped compounds,
e.g., a Haldane gap compound~\cite{Goto}. 
The peak behavior is explained by an additional nuclear spin-lattice relaxation 
induced by unpaired moments around defects or at crystal imperfections in the spin-gapped state. 
Using an electron spin-spin correlation function with a decay time $\tau_{e}$, we have
\begin{equation}
1/T_{1}=c^{2}A^{2}{\tau_{e}\over {1+(2\pi\nu_{\mathrm{res}}\tau_{e})^{2}}},
\label{e.T1}
\end{equation}
where $c$ is the number of ``impurity" relaxation centers per unit volume, 
$A$ is a nuclear-electron coupling constant, and $\nu_{\mathrm{res}}$ is the NMR frequency.~\cite{Goto,McHenry,BPP}
In the spin-gapped state, the lifetime $\tau_{e}$ of the impurity moment increases with cooling
and then $\nu_{\mathrm{res}}\tau_{e}$ = 1 leads to a peak in 1/$T_{1}$ of eq. (3).  
The absence of the peak behavior in 1/$^{125}T_{1}$ and 1/$^{65}T_{1}$
may be due to the large shifts of the impurity-neighbor Cu and Te NMR frequencies 
and 
the slow fluctuations of large hyperfine coupling constants $A$ at the Te and Cu sites.
The 30 K peak of 1/$^{23}T_{1}$ is enhanced by Zn impurities,
whereas it is suppressed by Na deficiency.
One can consider that as the spin vacancy increases by Zn substitution, 
$c$ increases but $\tau_{e}$ is invariant, 
whereas   
as the holes are doped by Na deficiency,
 $c$ increases and $\tau_{e}$ largely changes.
The static magnetic response to the spin vacancy is similar to that of the doped holes,
while the dynamical response is different from each other.  
This difference may result from the hopping motion of the doped holes.
 
In conclusion, a finite covalency at the Te site in the Cu-O-Te-O-Cu bond
was evidenced by the $^{125}$Te Knight shift measurement  
of Na$_{2}$Cu$_{2}$TeO$_{6}$.
We succeeded in introducing a topochemical Na deficiency by soft chemical treatment.
It turned out that the spin dynamics of the doped holes through Na deficiency is 
different from that of the spin vacancy through Zn impurities. 
 
\section*{Acknowledgement}
We thank Mr. H. Ohta, Drs. T. Waki and C. Michioka for their experimental supports
and helpful discussions, and Prof. M. Matsumoto for his supports for 
the ICP-AES experiments.   
This study was supported by a Grant-in-Aid 
for Science Research on Priority Area,
``Invention of anomalous quantum materials" from the Ministry of 
Education, Science, Sports and Culture of Japan (Grant No. 16076210).
% references

\end{document}